# ON THE FRACTAL PROPERTIES MICROACCELERATIONS

A.V. Sedelnikov

Samara State Aerospace University, Russia

443026, Russia, Samara, p/b 1253

e-mail: axe_backdraft@inbox.ru

In this paper the fractal property of the internal environment of space laboratory microaccelerations that occur. Changing the size of the space lab leads to the fact that the dependence of microaccelerations from time to time has the property similar to the self-affinity of fractal functions. With the help of microaccelerations, based on the model of the real part of the fractal Weierstrass-Mandelbrot function is proposed to form the inertial-mass characteristics of laboratory space with a given level of microaccelerations.

Key words: microaccelerations, self-affinity, fractal Weierstrass-Mandelbrot function.

## 1. Introduction

Modern space materials is growing rapidly at present. The unique conditions of space can be realized in space, such processes are not possible on Earth. Weightlessness one of these unique properties. The absence of convective motions such as helping to get unique content. For successful implementation of technological processes require large solar panels that provide electric power equipment. Efficient operation of these batteries is not possible without their orientation to the sun. For this purpose, the orientation engines. When you switch engines create a moment around the center of mass of the space lab, turn it and provide the specified orientation. For example, the draft of the Soviet space laboratory "NIKA-T" (1989) cosine of the angle between the normal to the panel and the direction of the sun should not be below 0.9. Active orientation of the laboratory with the help of engines leads to excitation of solar panels oscillations. These oscillations shake itself, and the laboratory. Due to this, in the internal environment created by the field laboratory microaccelerations. Accelerations can disrupt favorable conditions for some processes. Such processes are known as gravity-sensitive. Study, modeling and control of microaccelerations is an important problem of space materials. The high microaccelerations level on modern spacecraft significantly hinders the development of this advanced branch of science.

**2. Task statement and analysis of the problem**

The success of the gravity-sensitive process causes restrictions on the modulus of accelerations. So for some processes, these restrictions are as follows:

$$\begin{cases} |\vec{w}| \leq 10^{-6} g & \text{for frequencies } f < 0{,}1 \text{ Hz;} \\ |\vec{w}| \leq 10^{-5} \cdot f \cdot g & \text{for frequencies } 0{,}1 \leq f \leq 100 \text{ Hz;} \\ |\vec{w}| \leq 10^{-3} g & \text{for frequencies } f > 100 \text{ Hz,} \end{cases}$$

where f – oscillation frequency.

Therefore, one of the most important characteristics of modern laboratory space is the microaccelerations level. There have been attempts to create a laboratory without large flexible elements. In section 1.6 [1] described a series of spacecraft, "Foton." These spacecrafts run from 1985 to 2006. Power to them was carried out by the battery. Low microaccelerations levels was achieved due to the absence of the main source of microaccelerations - large fluctuations of the elastic elements. However, the spacecraft had a serious shortcoming. Their term of active existence does not exceed 18 days. In such circumstances it is impossible for lengthy processes. For example, the draft process of the spacecraft "NIKA-T" lifetime was 120 days. Spacecraft "Foton" could be a prototype space mini-factory of the future. Therefore, the new spacecraft of this series, "Foton-M4" is designed with solar panels. Seriously out of date and the idea of disposable spacecraft. Continuous launches and landings landing capsule will cause irreparable damage to the ecology of the planet, if used in the production space. A new concept of reusable space laboratory servicing on board the International Space Station is a project of "OKA-T" This spacecraft is detached from the International Space Station and goes offline for the duration of the flight process. Then again docked for loading of new raw materials. This process is repeated periodically. However, the problem of providing a given microaccelerations level has not been solved. The structural layout circuits labs are designed based on the needs of optimal placement of equipment, and then evaluated the microaccelerations level. If it is too high, then begins the process of repackaging. It requires a lot of time and resources. In this paper, we propose to go a different way. To do this, you should use a fractal property of the accelerations, which will be discussed below. The problem of selection of parameters space laboratory to a predetermined acceptable microaccelerations level. This approach is new and allows much faster to achieve the goal of delivered acceptable microaccelerations level.

## 3. Research and results

Consider the real part of the fractal Weierstrass-Mandelbrot function [2]:

$$\operatorname{Re} W(t) = C(t) = \sum_{n=-\infty}^{\infty} \frac{1-\cos b^n t}{b^{(2-D)n}}, \qquad (1)$$

where $D$ – fractal dimension, a $b$ – scale parameter of the fractal Weierstrass-Mandelbrot function. It satisfies the criteria of homogeneity [3]:

$$C(bt) = b^{2-D} C(t)$$

It follows that when $t$ is replaced by $b^4 t$, and $C(t)$ at $b^{4(2-D)}C(t)$ the appearance of the function remains unchanged. Thus, C (t) is self-affinity function.

We consider the problem of microaccelerations estimating of the internal environment of space lab [1]. In the physical formulation of this problem microacceleration generated mainly due to the natural vibrations of large flexible elements (eg solar panels). When you switch the orientation engines, they begin to vibrate and create microacceleration in the internal environment of space lab [4]. To explain the ideas we choose the simplest scheme of laboratory space in the form of a central spherical body and a resilient element. This element can be rigidly fixed to the chassis beam. We construct a dynamic microaccelerations after engine shutdown guidance for the following parameters of the circuit space of the laboratory:

Tab. 1

| Parameter | Dimension | Value |
|---|---|---|
| Moment of the orientation engine | N m | 6 |
| Mass of the central body | kg | 6000 |
| Mass per unit length of flexible element | kg/m | 20 |
| Length of flexible element | m | 6 |
| Coordinate of the attachment point $s$ | m | 1 |
| Central body moment of inertia | kg m$^2$ | 3375 |

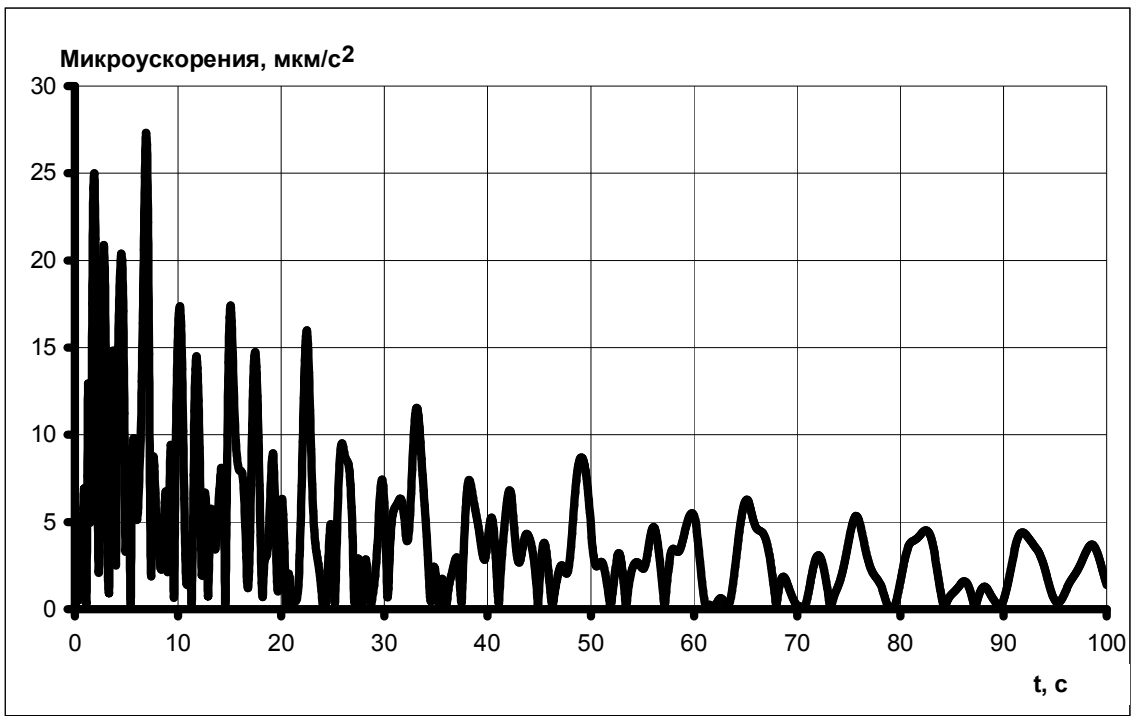

Fig. 1. The dynamics of accelerations during the first 100 s after switching off the engine orientation

We took into account the first six forms of natural vibrations, and the logarithmic decrement of vibrations was taken to be 0.1 [5]. In this case the natural frequency when the console consolidation expressed by the formula [6]:

$$\omega_i = \frac{\eta_i^2}{l^2}\sqrt{\frac{EI}{\mu}},$$

where $\eta_i$ – i-th root of the equation $\cos\eta_i \, ch\eta_i + 1 = 0$, $i$ – number of vibration modes, $l$ – length, $EI$ – flexural stiffness, a $\mu$ – mass per unit length of flexible element. For a given material per unit mass and stiffness remain constant. The roots $\eta_i$ are also constant. Thus, the frequency will be a function of beam length. So doubling the length of a decrease would lead to a fourfold increase in the frequency or four-fold decrease in the oscillation period. At the same time change the amplitude of the microaccelerations. Beam of shorter length and weight will create smaller microacceleration at a fixed central body space laboratory. At each point of the beam acts tangential force of inertia:

$$d\Phi_\tau = \mu \frac{M}{I}(x+s)\,dx,$$

where $M$ – moment of orientation engine, $s$ – projection of the radius vector of the point of attachment of the flexible element relative to the center of mass space laboratory to the axis x (Fig. 2), $I = I_0 + (1/3\,\mu l^3 + \mu l s^2)$ – moment of space laboratory inertia. In this case the beam is uniform. $I_0$ – moment of inertia of the central body.

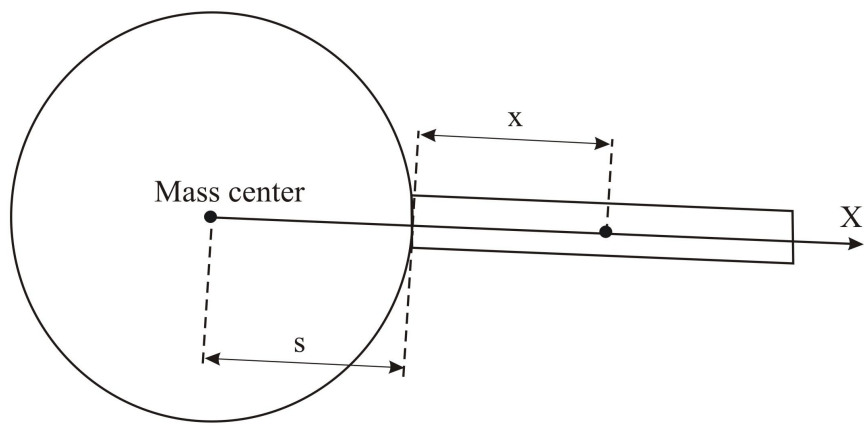

Fig. 2. Circuit of space lab

Then the total inertial force of the elastic element:

$$\Phi_\tau(l) = \mu \frac{M}{2I}(x+s)^2 = \frac{\mu}{2} \frac{M(l+s)^2}{I_0 + \frac{1}{3}\mu l^3 + \mu l s^2}.$$

When reducing the length of the beam is twice the formula is as follows:

$$\Phi_\tau(l/2) = \frac{\mu}{2} \frac{M(l/2+s)^2}{I_0 + \frac{1}{24}\mu l^3 + \mu \frac{l}{2} s^2}.$$

The substitution table data corresponding to the design of spacecraft such as "NIKA-T" [1] shows that in the first case, the force of inertia is approximately 0.596 N, while the second - 0.266 N. Thus, it has decreased more than twice. Without taking into account the energy loss in the attachment site of an elastic element, this means reducing the amplitude of the forces of reaction termination. This force is transmitted to the spacecraft body and serves as a source of microaccelerations, because it provides a moment about its center of mass.

For the initial amplitude of the microaccelerations can reduce the weight and size of the spacecraft so that the angular acceleration of the tangential forces of inertia have been the same.

$$\varepsilon(l) = \frac{M[\Phi_\tau(l)]}{I} = \frac{\Phi_\tau(l)s}{I} = \frac{\mu}{2} \frac{M(l+s)^2 s}{(I_0 + 1/3\mu l^3 + \mu l s^2)^2}, \qquad (2)$$

$$\varepsilon(l/2) = \frac{M[\Phi_\tau(l/2)]}{I_1} = \frac{\Phi_\tau(l)s_1}{I_1} = \frac{\mu}{2} \frac{M(l/2+s_1)^2 s_1}{(I_{10} + 1/24\mu l^3 + \mu l s_1^2/2)^2}. \qquad (3)$$

The moment of inertia of the central body is calculated by the formula: $I_{i0} = \frac{m_i R_i^2}{4}$, where $m_i$ and $R_i$ – the mass and radius of the central body for the i-th iteration. In the construction of a fractal model based on the accelerations Weierstrass-Mandelbrot function in [6] proposed to introduce a generic parameter z, which characterizes the mass fraction of the flexible elements in the total mass of the space lab:

$$z = \frac{100}{m_0 + \sum_{i=1}^{N}\mu_i l_i} \sum_{i=1}^{N}\mu_i l_i,$$

where $m_0$ – central body mass, a $N$ – number of flexible elements of the space lab. For this case:

$$z = \frac{100\,\mu l}{m_0 + \mu l}.$$

Similarly, (2) and (3) can be written expressions for the angular accelerations in further reducing the length of the flexible element:

$$l_i = \left(\frac{1}{2}\right)^i l. \qquad (4)$$

Then the equation:

$$\varepsilon(l) = \varepsilon(l/2) = \varepsilon_i$$

would guarantee the equality of the amplitudes of the forces of reaction in the clamped and, consequently, the amplitude values of the microaccelerations. Equation (4) provides a schedule compression dynamics microaccelerations from time to time due to a fourfold decrease in the period. Studies carried out for the space laboratory data (Tab. 1) show that this generalized parameter z also varies by about the same size (Fig. 3)

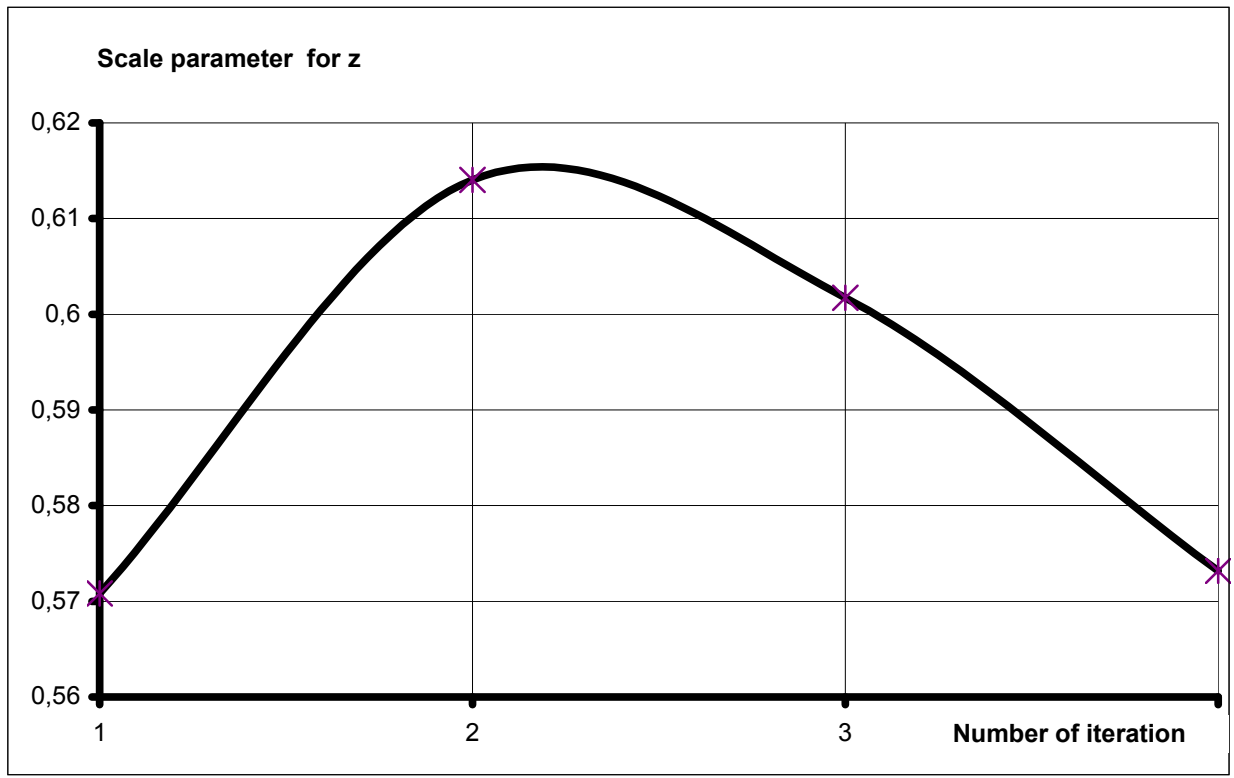

Fig. 3. Large-scale change parameter values of z for different iterations

Moreover, the variance of this value is 0.000455. Thus, it can be argued that as the length of the elastic element of the formula (4) in the selected simplest scheme is the reduction of the generalized parameter z by the approximate formula:

$$z_i = (0{,}59)^i \, z \tag{5}$$

But this means that the dynamics of microaccelerations in time has the property samoafinnosti, and z is the scale parameter, like b in the case of the Weierstrass-Mandelbrot function. The calculations show that the formula (5) is valid for large values of i. However, the practical value of the interest i > 6 is not present. o for the first iteration, using (5) we obtain the dependence of accelerations, shown in Fig. 4. After the second in Fig. 54.

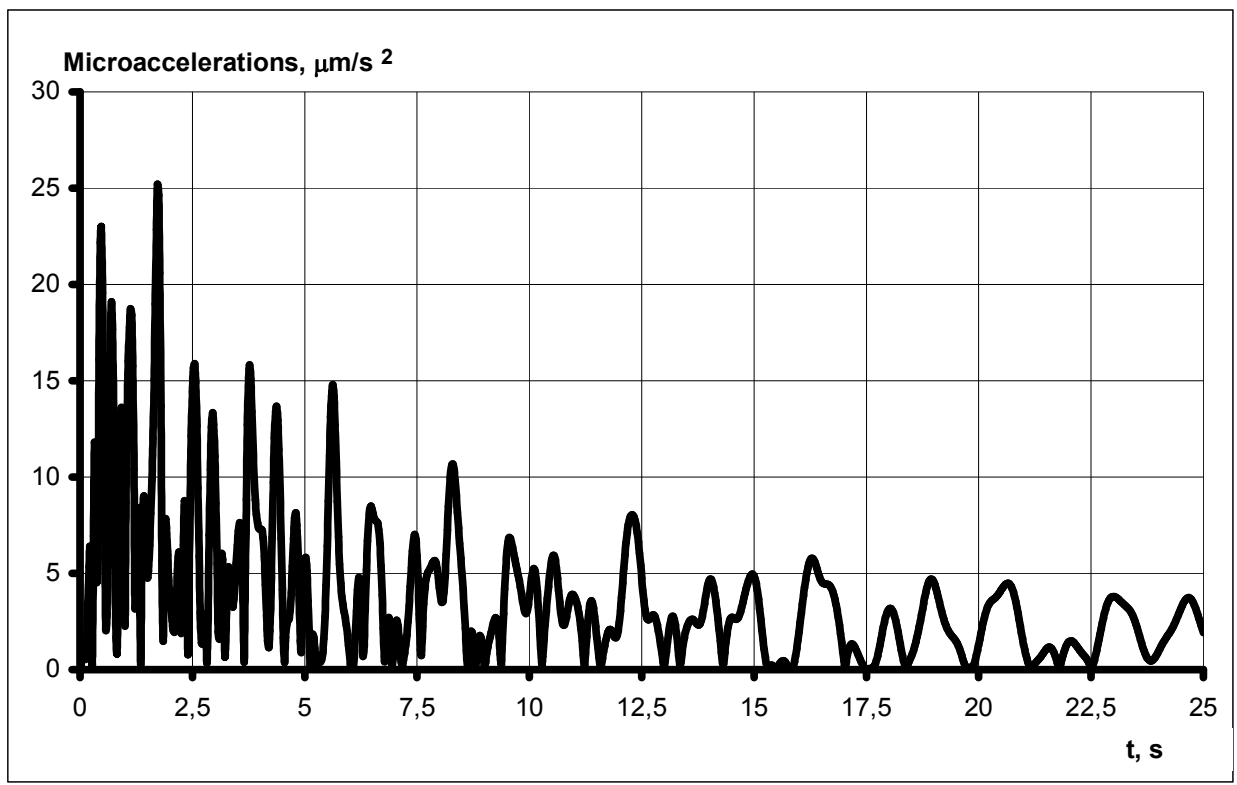

Fig. 4. The dependence of the microaccelerations on the time after the first iteration

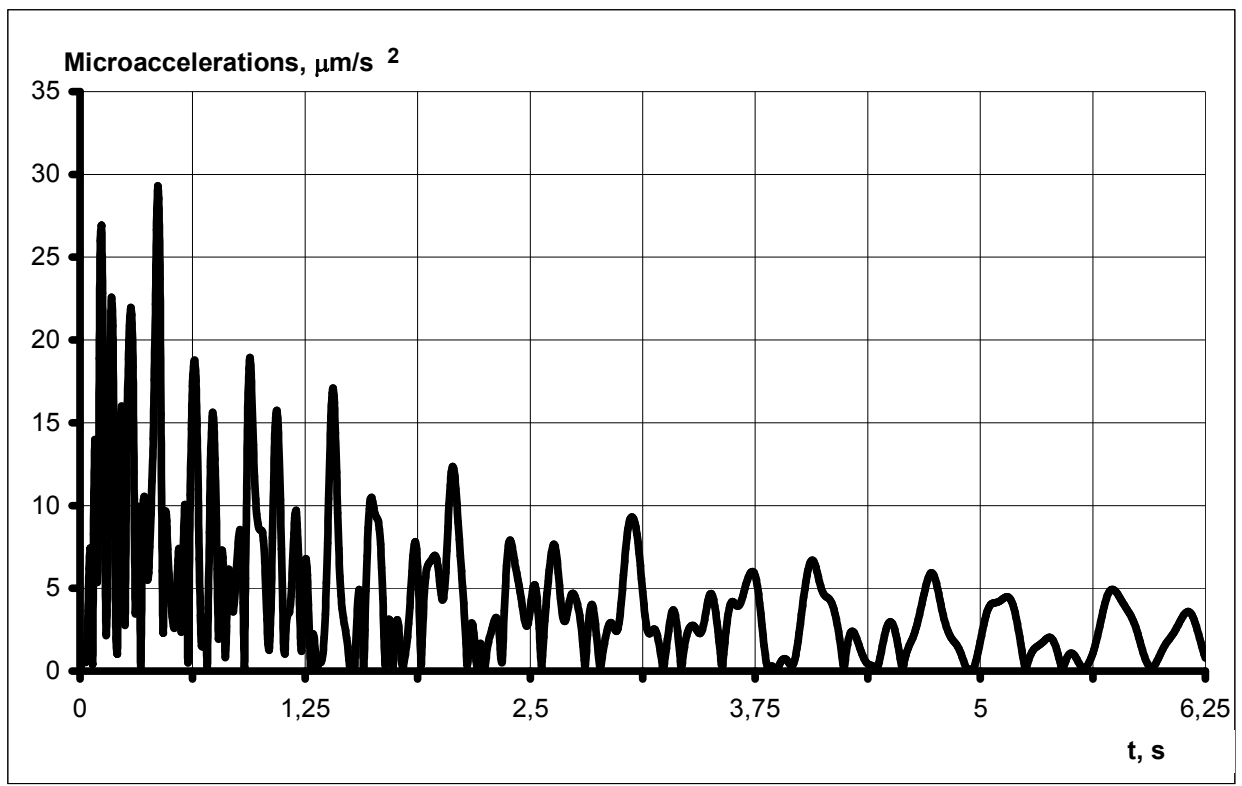

Fig. 5. The dependence of the microaccelerations of the time after the second iteration

The appearance of features in Fig. 1, 4 and 5 is somewhat different from each other, since the scale parameter for z at each step is different. However, with a certain accuracy can be argued that this relationship retains its shape when scaled in some way the most space laboratory.

**4. Summary and Conclusions**

Public property can be used for practical estimates of microaccelerations. The main characteristic of modern laboratory space is precisely the microaccelerations level. It determines the gravity-sensitive processes that have been successfully carried out on board the space lab. Therefore, before setting the allowable dynamic microaccelerations, can be achieved by scaling the laboratory to get the whole set of values of the generalized parameter z, which satisfies the given constraints on the microacceleration. Next, with z can be generated inertial-mass characteristics of the actual laboratory [7], corresponding to a predetermined dynamic microaccelerations. A number of studies of the author, for example [8-9] shows that for the fractal model microaccelerations can use the Weierstrass-Mandelbrot (1). In [9] carried out the identification of the parameters of the function of the Weierstrass-Mandelbrot factors leading to the field of microaccelerations in the internal environment of space laboratory. This model, in combination with the described property will appreciate the dynamics of the microaccelerations in the area of the proposed placement of technological equipment. At the same time can be optimized design and integration design lab to minimize the microaccelerations of the module.

For a variety of space laboratory scale parameter for z is somewhat different. As a result of these projects were analyzed by two Russian space laboratory "NIKA-T" and "OKA-T".